# An Optimum Multilevel Dynamic Round Robin Scheduling Algorithm

Neetu Goel

Research Scholar

TEERTHANKER MAHAVEER UNIVERSITY

e-mail:neetugoel1@rediffmail.com

Dr. R.B. Garg

Professor

Tecnia Institute of Advanced Studies

(Affiliated to GGSIP University, Delhi)

email: garg1943@gmail.com

**ABSTRACT**

The main objective of this paper is to improve the Round Robin scheduling algorithm using the dynamic time slice concept. CPU scheduling becomes very important in accomplishing the operating system (OS) design goals. The intention should be allowed as many as possible running processes at all time in order to make best use of CPU. CPU scheduling has strong effect on resource utilization as well as overall performance of the system. Round Robin algorithm performs optimally in timeshared systems, but it is not suitable for soft real time systems, because it gives more number of context switches, larger waiting time and larger response time. In this paper, a new CPU scheduling algorithm called *An Optimum Multilevel Dynamic Round Robin Scheduling Algorithm* is proposed, which calculates intelligent time slice and changes after every round of execution. The suggested algorithm was evaluated on some CPU scheduling objectives and it was observed that this algorithm gave good performance as compared to the other existing CPU scheduling algorithms.

**INTRODUCTION**

In a single-processor system, only one process can run at a time; any others must wait until the CPU is free and can be rescheduled. The objective of multiprogramming is to have some process running at all times, to maximize CPU utilization [1]. Scheduling is a fundamental operating-system function. Almost all computer resources are scheduled before use. The CPU is, of course, one of the primary computer resources. Thus, its scheduling is central to operating-system design. CPU scheduling determines which processes run when there are multiple run-able processes. CPU scheduling is important because it can have a big effect on resource utilization and the overall performance of the system [2].

OS may feature up to 3 distinct types of schedulers: a long term scheduler (also known as an admission scheduler or high level scheduler), a mid-term or medium-term scheduler and a short-term scheduler (also known as a dispatcher or CPU scheduler). In Round Robin (RR) every process has equal priority and is given a time quantum or time slice after which the process is preempted. Although RR gives improved response time and uses shared resources efficiently. Its limitations are larger waiting time, undesirable overhead and larger turnaround time for processes with variable CPU bursts due to use of static time quantum This motivates us to implement RR algorithm with sorted remaining burst time with dynamic time quantum concept. Another concept employed in this algorithm is the use of more than one cycle instead of a single Round Robin.

According to Seltzer, M P. Chen and J Outerhout 1990[3], the last thirty years have seen an enormous amount of research in the area of disk scheduling algorithm. The core objective has been to develop scheduling algorithms suited for certain goals sometimes with provable properties.

According to Sabrina, F.C.D, Nguyen, S.Jha, D. Platt and F. Safaei[2] Scheduling is a fundamental operating system function. Almost all computer resources are scheduled before use. The CPU is of course one of the primary resources. Thus its scheduling is central to Operating system design. CPU scheduling determines which process run when there are multiple run able processes CPU scheduling is important because it can have a big effect on resources utilization and overall performance of the system.

**SCHEDULING CRITERIA**

Different CPU scheduling algorithms have different properties, and the choice of a particular algorithm may favor one class of processes over another. In choosing which algorithm to use in a particular situation, we must consider the properties of the various algorithms. Many criteria have been suggested for comparing CPU scheduling algorithms. Which characteristics are used for comparison can make a substantial difference in which algorithm is judged to be best. The criteria include the following:

1. **Utilization/Efficiency**: keep the CPU busy 100% of the time with useful work
2. **Throughput:** maximize the number of jobs processed per hour.

3. **Turnaround time**: from the time of submission to the time of completion, minimize the time batch users must wait for output
4. **Waiting time**: Sum of times spent in ready queue - Minimize this
5. **Response Time**: time from submission till the first response is produced, minimize response time for interactive users
6. **Fairness**: make sure each process gets a fair share of the CPU

**ORGANIZATION OF THE PAPER**

The paper is divided into four sections. *Section I* gives a brief introduction on the various aspects of the scheduling algorithms, the approach to the current paper and the motivational factors leading to this improvement. *Section II* presents the proposed algorithm and illustration of our proposed new algorithm (OMDRR). In *Section III*, an experimental analysis and Result of our algorithm (OMDRR) and its comparison with the static RR algorithm. Conclusion is presented in *Section IV* followed up by the references used.

**II. PROPOSED ALGORITHM**

The early the shorter processes are removed from the ready queue, the better the turnaround time and the waiting time. So in our algorithm, the shorter processes are given more time quantum so that they can finish their execution earlier. Here shorter processes are defined as the processes having less assumed CPU burst time than the previous process. Performance of RR algorithm solely depends upon the size of time quantum. If it is very small, it causes too many context switches. If it is very large, the algorithm degenerates to FCFS. So our algorithm solves this problem by taking dynamic intelligent time quantum where the time quantum is repeatedly adjusted according to the shortness component.

### A. Our Proposed Algorithm

In our algorithm, combines the working principle of fundamental scheduling algorithms. Dynamically Time Slice (DTS) is calculated which allocates different time quantum to each process based on priority, shortest CPU burst time and context switch avoidance time.

**Step 1:** Shuffle the processes in ascending order in the ready queue such that the head of the ready queue contains the lowest burst time.

**Step 2:** If one or more process has equal burst time then
    {
        Allocate the CPU to the processes according to First Come basis.
    }

**Step 3:** Assign the time quantum and apply for each process say TQ=k.

**Step 4:** IF (burst time of the process < TQ)
    {
        Allocate the CPU to that process till it terminates.
    }
    ELSE IF (Remaining burst time of the process < TQ/2)
    {
        Allocate the CPU again to that process till it terminates.
    }
    ELSE
    {

        (i) The process will occupy the CPU till the time quantum and it is added to the ready queue in ascending order for the next round of execution.

        (ii) TQ= TQ *2

        (iii) K=TQ

        (iv) Goto Step 3

        }

**III EXPERIMENTAL ANALYSIS**

Five processes have been defined with CPU burst time, these five processes are scheduled in round robin and also in the proposed algorithm. The context switch, average waiting time, average turn around time has been calculated and the results were compared. For doing this we have carry a number of experiments but here I will discuss only two experiments because we assured results analysis is remain unchanged. This paper presents two types of problem analysis where first one is without arrival time and second one is with arrival time.

**Experiment A:**

In this we have to consider the processes only with CPU burst time and also let round robin quantum =5

According to simple RR scheduling:

| Process Id | CPU Burst Time (ms) |
| --- | --- |
| P1 | 22 |
| P2 | 18 |
| P3 | 9 |
| P4 | 10 |
| P5 | 5 |

RR quantum=5

According to the simple RR algorithm:

**Gantt chart:**

| P1 | P2 | P3 | P4 | P5 | P1 | P2 | P3 | P4 | P1 | P2 | P1 | P2 | P1 |
| --- | --- | --- | --- | --- | --- | --- | --- | --- | --- | --- | --- | --- | --- |

No. of context switches =13

Average waiting time=34 ms

Average turnaround time= 46.8 ms

**According to proposed algorithm:-**

Gantt chart:

| P1 | P2 | P3 | P4 | P5 | P3 | P4 | P1 | P2 | P1 |

No of context switches = 9

Average waiting time =28.6 ms

Average turn around time = 41.4 ms

**Experiment B:**

In this we have to consider the processes along with CPU burst time and process arrival time and also let round robin quantum =3

According to simple RR scheduling:

| Process Id | Arrival Time | CPU Burst Time(ms) |
|---|---|---|
| P1 | 0 | 4 |
| P2 | 2.4 | 7 |
| P3 | 5.1 | 5 |
| P4 | 6.2 | 8 |
| P5 | 8.019 | 9 |

According to the simple RR algorithm:

Gantt chart:

| P1 | P2 | P3 | P4 | P5 | P1 | P2 | P3 | P4 | P5 | P2 | P4 | P5 |

No of context switches = 12

Average waiting time =19 ms

Average turn around time = 25.6 ms

According to proposed algorithm:-

Gantt chart:

| P1 | P2 | P3 | P4 | P5 | P3 | P2 | P4 | P5 |

No of context switches = 8

Average waiting time =14.2 ms

Average turn around time = 20.8 ms

We can see from the above experiment context switch, average waiting time and average turnaround time both are reduced by using our proposed algorithm. The reduction of context switch, average waiting time and average turnaround time shows maximum CPU utilization and minimum response time. We observed that proposed algorithm much more efficient as compared to simple RR algorithm.

**CONCLUSION**

A comparative study of simple RR algorithm and proposed one is made. It is concluded that the proposed algorithm is superior as it has less waiting response time, usually less pre-emption and context switching thereby reducing the overhead and saving of memory space. Future work can be based on this algorithm modified and implemented for hard real time system where hard deadline systems require partial outputs to prevent catastrophic events.